\documentclass[12pt,a4paper,english,nofootinbib]{revtex4}
\usepackage{lmodern}
\usepackage{lmodern}

\usepackage[T1]{fontenc}
\usepackage[latin9]{inputenc}
\usepackage{babel}
\usepackage{textcomp}
\usepackage{mathrsfs}
\usepackage{amsmath}
\usepackage{amssymb}
\usepackage{esint}
\usepackage[unicode=true,pdfusetitle,
 bookmarks=true,bookmarksnumbered=false,bookmarksopen=false,
 breaklinks=false,pdfborder={0 0 1},backref=false,colorlinks=false]
 {hyperref}

\makeatletter

\@ifundefined{textcolor}{}
{%
 \definecolor{BLACK}{gray}{0}
 \definecolor{WHITE}{gray}{1}
 \definecolor{RED}{rgb}{1,0,0}
 \definecolor{GREEN}{rgb}{0,1,0}
 \definecolor{BLUE}{rgb}{0,0,1}
 \definecolor{CYAN}{cmyk}{1,0,0,0}
 \definecolor{MAGENTA}{cmyk}{0,1,0,0}
 \definecolor{YELLOW}{cmyk}{0,0,1,0}
}

\usepackage{babel}
\usepackage{babel}
\usepackage{babel}
\usepackage{babel}
\usepackage{babel}
\usepackage{babel}
\usepackage{babel}
\usepackage{babel}
\usepackage{babel}
\usepackage{babel}
\usepackage{babel}
\usepackage{babel}
\usepackage{babel}
\usepackage{babel}

\usepackage{babel}
\usepackage{graphicx}
\def\b{\begin{equation}}
\def\e{\end{equation}}

\@ifundefined{textcolor}{}{%
 \definecolor{BLACK}{gray}{0}
 \definecolor{WHITE}{gray}{1}
 \definecolor{RED}{rgb}{1,0,0}
 \definecolor{GREEN}{rgb}{0,1,0}
 \definecolor{BLUE}{rgb}{0,0,1}
 \definecolor{CYAN}{cmyk}{1,0,0,0}
 \definecolor{MAGENTA}{cmyk}{0,1,0,0}
 \definecolor{YELLOW}{cmyk}{0,0,1,0}
 }

\usepackage{latexsym}\usepackage{bm}

\makeatother

\begin{document}
\title{{\normalsize{}{}{}NON-STATIONARY ENERGY IN GENERAL RELATIVITY}}
\author{{\normalsize{}{}{}{}{}{}{}{}{}{}Emel Altas}}
\email{emelaltas@kmu.edu.tr}

\affiliation{Department of Physics,\\
 Karamanoglu Mehmetbey University, 70100, Karaman, Turkey}
\author{{\normalsize{}{}{}{}{}{}{}{}{}{}Bayram Tekin}}
\email{btekin@metu.edu.tr}

\affiliation{Department of Physics,\\
 Middle East Technical University, 06800, Ankara, Turkey}

\maketitle

\date{{\normalsize{}{}{}{{}{}{}{}\today}}}

Using the time evolution equations of (cosmological) General Relativity
in the first order Fischer-Marsden form, we construct an integral
that measures the amount of non-stationary energy on a given spacelike
hypersurface in $D$ dimensions. The integral vanishes for stationary
spacetimes; and with a further assumption, reduces to Dain's invariant
on the boundary of the hypersurface which is defined with the Einstein
constraints and a fourth order equation defining approximate Killing
symmetries. 

\section{{\normalsize{}{}{}{}{}{}{}{}{}{}Introduction}}

Dain \cite{Dain} constructed a geometric invariant that measures
the \textit{non-stationary} energy for an asymptotically flat hypersurface
in 3+1 dimensions for the case of time-symmetric initial data which,
for vacuum, is an invariant that quantifies the total energy of the
gravitational radiation. So this invariant is a component of the total
ADM energy \cite{ADM} assigned to an asymptotically flat hypersurface.
That construction was extended to the time-non symmetric case recently
in \cite{Kroon}. To give an example of how useful such a geometric
invariant can be when constructing initial data for the gravitational
field, let us recall the first observation of the merger of two black
holes \cite{merger}. According to this observation, two initial black
holes with masses (approximately) 36$\textup{M}_{\odot}$ and 29$\textup{M}_{\odot}$
merged to produce a single stationary black hole of mass 62$\textup{M}_{\odot}$
plus gravitational radiation of total energy equivalent to 3$\textup{M}_{\odot}$.
Assuming this system to be isolated in an asymptotically flat spacetime,
the total initial ADM energy of 65$\textup{M}_{\odot}$ is certainly
conserved. But this total ADM energy of the initial data needs a refinement
as it clearly has a non-stationary part equal to 3$\textup{M}_{\odot}$.
The important question is to identify this non-stationary energy in
the initial data. 

Dain's construction and its extension to the non-time symmetric case
by Kroon and Williams \cite{Kroon} are based on several earlier crucial
works one of which is the Killing initial data (KID) concept of Moncrief
\cite{Moncrief} and Beig-Chru\'{s}ciel \cite{Beig}; and a fourth
order operator defined by Bartnik \cite{Bartnik}. Of course all of
the discussion is related to the Cauchy problem in General Relativity
and the related issue of constructing initial data for the time evolution
equations. Here by using the time-evolution equations, in the form
given by Fischer and Marsden \cite{FisMar72}, we construct a new
representation of the non-stationary energy in generic $D$ dimensional
spacetimes with or without a cosmological constant. 

The outline of the paper is as follows: in section II we briefly summarize
Dain's construction using the constraints and present a new approach
using the evolution equations. In section III we give the details
of the relevant computations in $D$ dimensions . The Appendix is
devoted to the ADM decomposition. 

\section{{\normalsize{}{}{}Dain's invariant in brief and a new formulation}}

Leaving the details of the construction to the next section, let us
first briefly summarize the ingredients needed to define Dain's invariant
on a spacelike hypersurface $\Sigma$ of the spacetime $\mathscr{M}=\mathbb{R}\times\Sigma$.
Then we shall discuss our new formulation via the evolution equations.

The initial data on the hypersurface is defined by the Riemannian
metric $\gamma_{ij}$ and the extrinsic curvature $K_{ij}$ in local
coordinates. Denoting $D_{i}$ to be the covariant derivative compatible
with $\gamma_{ij}$ and assuming the usual ADM decomposition of the
spacetime metric $g_{\mu\nu}$, the line element reads 
\begin{equation}
ds^{2}=(N_{i}N^{i}-N^{2})dt^{2}+2N_{i}dtdx^{i}+\gamma_{ij}dx^{i}dx^{j},\label{ADMdecompositionofmetric}
\end{equation}
while the extrinsic curvature becomes \footnote{Our definition of the extrinsic curvature is as follows: given $(X,Y)$
two vectors on the tangent space $T_{p}\Sigma$ and $n$ be the unit
normal to $\Sigma$, then $K(X,Y):=g(\nabla_{X}n,Y)$ with $\nabla$
being the covariant derivative compatible with the spacetime metric
$g$. } 
\begin{equation}
K_{ij}=\frac{1}{2N}\left(\dot{\gamma}_{ij}-D_{i}N_{j}-D_{j}N_{i}\right),
\end{equation}
with the lapse function $N=N(t,x^{i})$ and the shift vector $N^{i}=N^{i}(t,x^{i})$.
The spatial indices can be raised and lowered with the $D-1$ dimensional
spatial metric $\gamma$; over dot denotes the derivative with respect
to $t$, and the Latin letters are used for the spatial dimensions,
$i,j,k,...=1,2,3,...D-1$, whereas the Greek letters are used to denote
the spacetime dimensions, $\mu,\nu,\rho,...=0,1,2,3,...D-1$. All
the relevant details of the ADM decomposition are given in the Appendix.

Under the above decomposition of spacetime, the $D$-dimensional Einstein
equations 
\begin{equation}
R_{\mu\nu}-\frac{1}{2}Rg_{\mu\nu}+\Lambda g_{\mu\nu}=\kappa T_{\mu\nu}
\end{equation}
yield the Hamiltonian and momentum constraints on the hypersurface
$\Sigma$ as 
\begin{eqnarray}
 &  & \Phi_{0}(\gamma,K):=-{}^{\Sigma}R-K^{2}+K_{ij}^{2}+2\Lambda-2\kappa T_{nn}=0,\nonumber \\
 &  & \Phi_{i}(\gamma,K):=-2D_{k}K_{i}^{k}+2D_{i}K-2\kappa T_{ni}=0,~~~~~~~~~~~~~~\label{Einstein_c}
\end{eqnarray}
where $K:=\gamma^{ij}K_{ij}$ and $K_{ij}^{2}:=K^{ij}K_{ij}$. From
now on we shall work in vacuum, hence $T_{\mu\nu}=0$. Denoting $\Phi(\gamma,K)$
to be the constraint covector with components ($\Phi_{0},\Phi_{i}$)
and $D\Phi(\gamma,K)$ to be its linearization about a given solution
$(\gamma,K)$ to the constraints and $D\Phi^{*}(\gamma,K)$ to be
the formal adjoint map, then following Bartnik \cite{Bartnik}, one
defines another operator ${\cal {P}}$: 
\begin{equation}
{\cal {P}}:=D\Phi(\gamma,K)\circ\begin{pmatrix}1 & 0\\
0 & -D^{m}
\end{pmatrix}.
\end{equation}
The reason why we need this operator will be clear below. Using the
formal adjoint ${\cal {P}}^{*}$ of Bartnik's operator, Dain \cite{Dain}
defines the following integral over the hypersurface 
\begin{equation}
\mathscr{I}(N,N^{i}):=\intop_{\Sigma}dV~{\cal {P}}^{*}\begin{pmatrix}N\\
N^{k}
\end{pmatrix}\cdot{\cal {P}}^{*}\begin{pmatrix}N\\
N^{k}
\end{pmatrix},\label{integralinvariant}
\end{equation}
where the multiplication is defined as 
\begin{equation}
\begin{pmatrix}N\\
N^{i}
\end{pmatrix}\cdot\begin{pmatrix}A\\
B_{i}
\end{pmatrix}:=NA+N^{i}B_{i}.
\end{equation}
The integral (\ref{integralinvariant}) is to be evaluated for specific
vectors $\xi:=(N,N^{i})$ that satisfy the fourth-order equation 
\begin{equation}
{\cal {P}}\circ{\cal {P}}^{*}\left(\xi\right)=0,\label{approx_KID}
\end{equation}
which Dain called the \textit{approximate Killing initial data} (KID)
equation. It is clear that if $\xi$ satisfies the lower derivative
equation ${\cal {P}}^{*}\left(\xi\right)=0$, then it also satisfies
(\ref{approx_KID}). Moreover, these particular solutions, together
with an assumption on their decay at infinity, also solve the KID
equations which are simply $D\Phi^{*}(\gamma,K)\left(\xi\right)=0$.
In fact this point is crucial but well-established: Moncrief \cite{Moncrief}
proved that $\xi$ is a spacetime Killing vector satisfying $\nabla_{\mu}\xi_{\nu}+\nabla_{\nu}\xi_{\mu}=0$
if and only if it satisfies the KID equations. Namely one has 
\begin{equation}
\nabla_{\mu}\xi_{\nu}+\nabla_{\nu}\xi_{\mu}=0\Leftrightarrow D\Phi^{*}(\gamma,K)\left(\xi\right)=0,
\end{equation}
with $(N,N^{i})$ being the projections off and onto the hypersurface
of the Killing vector field $\xi$. The physical picture is clear:
initial data on the hypersurface clearly encode the spacetime symmetries.
There have been rigorous works on the KIDs in \cite{BeiChr96,Chrusciel:2005pb,Beig}
which we shall employ in what follows.

Observe that for any Killing vector field $\text{\ensuremath{\mathscr{I}}}(N,N^{i})$
vanishes identically. So by design, Dain's invariant identically vanishes
for initial data with exact symmetries. Then Dain goes on to show
that for asymptotically flat spaces, for the case of approximate translational
KID's $\text{\ensuremath{\mathscr{I}}}(N,N^{i})$ can measure the
\textit{non-stationary energy } contained in the hypersurface $\Sigma$.
To simplify his calculations Dain considered the time symmetric initial
data ($K_{ij}=0$) in three spatial dimensions.There are two crucial
points to note about Dain's construction: firstly, one can show that
for any asymptotically flat three manifold, the approximate KID equation
has non-trivial solutions which are not KIDs; secondly, using integration
by parts, one can convert the volume integral (\ref{integralinvariant})
to a surface integral. We shall discuss these in the next section,
but let us first give another formulation of this invariant.

\subsection{Non-Stationary Energy via Time-evolution Equations}

In Dain's construction, as is clear from the above summary, time evolution
of the initial data has not played a role: in fact one only works
with the constraints on the hypersurface. This fact somewhat obscures
the interpretation of the proposed invariant as the non-stationary
energy contained in the initial data. In what follows, we propose
another formulation of this invariant with the help of the time evolution
equations which makes the interpretation clearer. For this purpose
let us consider the phase space variables to be the spatial metric
$\gamma_{ij}$ and the canonical momenta $\pi^{ij}$; the latter can
be found from the Einstein-Hilbert Lagrangian 
\begin{equation}
\text{\ensuremath{\mathscr{L}}}_{EH}=\frac{1}{\kappa}\sqrt{-g}(R-2\Lambda)=\frac{1}{\kappa}\sqrt{\gamma}N({}^{\Sigma}R+K_{ij}^{2}-K^{2}+\Lambda)+boundary~terms
\end{equation}
which are 
\begin{equation}
\pi^{ij}:=\frac{\delta\text{\ensuremath{\mathscr{L}}}_{EH}}{\delta\dot{\gamma_{ij}}}=\frac{1}{\kappa}\sqrt{\gamma}(K^{ij}-\gamma^{ij}K).
\end{equation}
Using the canonical momenta, it pays to recast the densitized versions
of the constraints (\ref{Einstein_c}) for $T_{\mu\nu}=0$ and setting
$\kappa=1$ as 
\begin{eqnarray}
 &  & \Phi_{0}(\gamma,\pi):=\sqrt{\gamma}\left(-^{\Sigma}R+2\Lambda\right)+G_{ijkl}\pi^{ij}\pi^{kl}=0,\nonumber \\
 &  & \Phi_{i}(\gamma,\pi):=-2\gamma_{ik}D_{j}\pi^{kj}=0,\label{einstein_c2}
\end{eqnarray}
where the \textit{DeWitt metric } \cite{DeWitt} $G_{ijkl}$ in $D$
dimensions reads 
\begin{equation}
G_{ijkl}=\frac{1}{2\sqrt{\gamma}}\left(\gamma_{ik}\gamma_{jl}+\gamma_{il}\gamma_{jk}-\frac{2}{D-2}\gamma_{ij}\gamma_{kl}\right).
\end{equation}
Ignoring the possible boundary terms, the ADM Hamiltonian density
turns out to be a sum of the constraints as 
\begin{equation}
{\cal {H}}=\int_{\Sigma}d^{D-1}x\thinspace\left\langle {\cal {N}},\Phi(\gamma,\pi)\right\rangle ,\label{hamilton}
\end{equation}
with ${\cal {N}}$ being the lapse-shift vector with components $(N,N^{i})$
which play the role of the Lagrange multipliers; and the angle-brackets
denote the usual contraction. Given an ${\cal {N}}$, the remaining
evolution equations can be written in a compact form (the Fischer-Marsden
form \cite{Fischer-Marsden}) as 
\begin{equation}
\frac{d}{dt}\begin{pmatrix}\gamma\\
\pi
\end{pmatrix}=J\circ D\Phi^{*}(\gamma,\pi)({\cal {N}}),\label{evolution}
\end{equation}
where the $J$ matrix reads 
\begin{equation}
J=\begin{pmatrix}0 & 1\\
-1 & 0
\end{pmatrix}.
\end{equation}
The reason why the formal adjoint of the linearized constraint map
$D\Phi^{*}(\gamma,\pi)$ appears in the time evolution can be seen
as follows: the Hamiltonian form of the Einstein-Hilbert action 
\begin{equation}
{\cal {S}}_{EH}\left[\gamma,\pi\right]=\intop dt\intop d^{D-1}x\left(\pi^{ij}\dot{\gamma}_{ij}-\left\langle {\cal {N}},\Phi(\gamma,\pi)\right\rangle \right),
\end{equation}
when varied about a background $\left(\gamma,\pi\right)$ gives 
\begin{equation}
D{\cal {S}}_{EH}\left[\gamma,\pi\right]=\intop dt\intop d^{D-1}x\left(\delta\pi^{ij}\dot{\gamma}_{ij}+\pi^{ij}\delta\dot{\gamma}_{ij}-\left\langle {\cal {N}},D\Phi(\gamma,\pi)\cdot\left(\delta\gamma,\delta\pi\right)\right\rangle \right).\label{actionvariation}
\end{equation}
Here the linearized form of the constraint map can be computed to
be 
\begin{equation}
D\Phi\begin{pmatrix}h_{ij}\\
p^{ij}
\end{pmatrix}=\begin{pmatrix}\sqrt{\gamma}\left(^{\varSigma}R{}^{ij}h_{ij}-D^{i}D^{j}h_{ij}+\triangle h\right)\\
-hG_{ijkl}\pi^{ij}\pi^{kl}+2G_{ijkl}p^{ij}\pi^{kl}+2G_{njkl}h_{im}\gamma^{mn}\pi^{ij}\pi^{kl}\\
\\
-2\gamma_{ik}D_{j}p^{kj}-\pi^{jk}\left(2D_{k}h_{ij}-D_{i}h_{jk}\right)
\end{pmatrix},\label{pppmatrix}
\end{equation}
where $\delta\gamma_{ij}:=h_{ij}$, $h:=\gamma^{ij}h_{ij}$, $\delta\pi^{ij}:=p^{ij}$
and $\triangle:=D_{k}D^{k}$. We have used the vanishing of the constraints
to simplify the expression. In (\ref{actionvariation}) using integration
by parts when necessary and dropping the boundary terms one arrives
at the desired result 
\begin{equation}
D{\cal {S}}_{EH}\left[\gamma,\pi\right]=\intop dt\intop d^{D-1}x\left(\delta\pi^{ij}\dot{\gamma}_{ij}-\dot{\pi}^{ij}\delta\gamma_{ij}-\left\langle \left(\delta\gamma,\delta\pi\right),D\Phi^{*}(\gamma,\pi)\cdot{\cal {N}}\right\rangle \right),\label{actionvariation3}
\end{equation}
where the adjoint constraint map appears in the process which reads
\begin{equation}
D\Phi^{*}\begin{pmatrix}N\\
N^{i}
\end{pmatrix}=\begin{pmatrix}\sqrt{\gamma}\left(^{\varSigma}R{}^{ij}-D^{i}D^{j}+\gamma^{ij}\triangle\right)N\\
-N\gamma^{ij}G_{klmn}\pi^{kl}\pi^{mn}+2NG_{klmn}\gamma^{ik}\pi^{jl}\pi^{mn}\\
+2\pi^{k(i}D_{k}N^{j)}-D_{k}(N^{k}\pi^{ij})\\
\\
2NG_{ijkl}\pi^{kl}+2D_{(i}N_{j)}
\end{pmatrix}.\label{formaladjointoflinearizedmap_pi}
\end{equation}
Setting the variation (\ref{actionvariation3}) to zero one obtains
the evolution equations (\ref{evolution}) or in more explicit form
one has
\begin{equation}
\frac{d\gamma_{ij}}{dt}=2NG_{ijkl}\pi^{kl}+2D_{(i}N_{j)},\label{ev1}
\end{equation}
and 
\begin{eqnarray}
 &  & \frac{d\pi^{ij}}{dt}=\sqrt{\gamma}\left(-^{\varSigma}R{}^{ij}+D^{i}D^{j}-\gamma^{ij}\triangle\right)N+N\gamma^{ij}G_{klmn}\pi^{kl}\pi^{mn}-2NG_{klmn}\gamma^{ik}\pi^{jl}\pi^{mn}\label{timederivativeofpi}\\
 &  & ~~~~~~~~-2\pi^{k(i}D_{k}N^{j)}+D_{k}(N^{k}\pi^{ij}).\nonumber 
\end{eqnarray}
Together with the constraints (\ref{einstein_c2}) these two tensor
equations constitute a set of constrained dynamical system for a \textit{given}
lapse-shift vector an $(N,N^{i})$. The constraints have a dual role:
they determine the viable initial data and also generate time evolution
of the initial data once the lapse-shift vector is chosen. As noted
above, if $D\Phi^{*}(\gamma,\pi)({\cal {N}})=0$, namely ${\cal {N}}=\xi$
is a Killing vector field then the time evolution is trivial. In particular
this would be the case for a stationary Killing vector.

Consider now an ${\cal {N}}$ which is \textit{not} a Killing vector,
which means $D\Phi^{*}(\gamma,\pi)({\cal {N}})\neq0$; and in particular
directly from the evolution equations we can find how much $D\Phi^{*}(\gamma,\pi)({\cal {N}})$
differs from zero (or how much a given ${\cal {N}}$ fails to be a
Killing vector) as 
\begin{equation}
D\Phi^{*}(\gamma,\pi)({\cal {N}})=J^{-1}\circ\frac{d}{dt}\begin{pmatrix}\gamma\\
\pi
\end{pmatrix}.
\end{equation}
To get a number from this matrix, first one should note that the units
of $\gamma$ and $\pi$ are different by a factor of $1/L$ and so
a naive approach of taking the "square" of this matrix does not
work. At this stage to remedy this, one needs the (adjoint) operator
of Bartnik that we have introduced above: so one has 
\begin{equation}
{\cal {P}}^{*}({\cal {N}}):=\begin{pmatrix}1 & 0\\
0 & D_{m}
\end{pmatrix}\circ D\Phi^{*}(\gamma,\pi)({\cal {N}})=\begin{pmatrix}1 & 0\\
0 & D_{m}
\end{pmatrix}\circ J^{-1}\circ\frac{d}{dt}\begin{pmatrix}\gamma\\
\pi
\end{pmatrix},
\end{equation}
which yields ${\cal {P}}^{*}({\cal {N}})=(-\dot{\pi},D_{m}\dot{\gamma})$.
Since $\pi$ is a tensor density to get a number out of this vector,
we further define 
\begin{equation}
\widetilde{{\cal {P}}}^{*}({\cal {N}}):=\begin{pmatrix}\gamma^{-1/2} & 0\\
0 & 1
\end{pmatrix}\circ{\cal {P}}^{*}({\cal {N}}).\label{formaladjoint_ptilde}
\end{equation}
Then the integral of $\widetilde{{\cal {P}}}^{*}({\cal {N}})\cdot\widetilde{{\cal {P}}}^{*}({\cal {N}})$
over the hypersurface yields 
\begin{equation}
\text{\ensuremath{\mathscr{I}}}({\cal {N}})=\intop_{\Sigma}dV\thinspace\widetilde{{\cal {P}}}^{*}({\cal {N}})\cdot\widetilde{{\cal {P}}}^{*}({\cal {N}})=\intop_{\Sigma}dV\thinspace\left(|D_{m}\dot{\gamma}_{ij}|^{2}+\frac{1}{\gamma}|\dot{\pi}^{ij}|^{2}\right),\label{bizimdenk}
\end{equation}
where $|D_{m}\dot{\gamma}_{ij}|^{2}:=\gamma^{mn}{\gamma}^{ij}{\gamma}^{kl}D_{m}\dot{\gamma}_{ik}D_{n}\dot{\gamma}_{jl}$
and $|\dot{\pi}^{ij}|^{2}:={\gamma}_{ij}{\gamma}_{kl}\dot{\pi}^{ik}\dot{\pi}^{jl}$.
This is another representation of Dain's invariant which explicitly
involves the time derivatives of the canonical fields. We have also
not assumed that the cosmological constant vanishes, hence our result
is valid for generic spacetimes. Note that this expression is valid
for any ${\cal {N}}$ which is not necessarily an approximate KID,
hence given a solution to the constraint equations and a choice of
the lapse-shift vector, one can compute this integral. But the volume
integral becomes a surface integral when ${\cal {N}}$ is an approximate
KID which is the case considered by Dain. Observe that by construction,
$\text{\ensuremath{\mathscr{I}}}({\cal {N}})$ is a non-negative number.
To get the explicit expression as a volume integral in terms of the
canonical fields and not their time derivatives, one should plug the
two evolution equations (\ref{ev1}) and (\ref{timederivativeofpi})
to (\ref{bizimdenk}). The resulting expression is 
\begin{eqnarray}
 &  & \text{\ensuremath{\mathscr{I}}}({\cal {N}})=\intop_{\Sigma}dV\Bigg\{|D_{m}V^{ij}|^{2}+{}^{\varSigma}R_{ij}^{2}N^{2}+(D_{i}D_{j}N)^{2}-2{}^{\varSigma}R^{ij}ND_{i}D_{j}N+2{}^{\varSigma}RN\triangle N\nonumber \\
 &  & ~~~~~~~~~~~~+(D-3)\triangle N\triangle N+2Q\triangle N+Q_{ij}^{2}+2{}^{\varSigma}R_{ij}NQ^{ij}-2Q^{ij}D_{i}D_{j}N\nonumber \\
 &  & ~~~~~~~~~~~~~+4D_{m}D_{(i}N_{j)}D^{m}D^{(i}N^{j)}+4D_{m}D_{i}N_{j}D^{m}V^{ij}\Bigg\},\label{esasintegral}
\end{eqnarray}
where 
\begin{equation}
V^{ij}:=\frac{2N}{\sqrt{\gamma}}\left(\pi^{ij}-\frac{1}{D-2}\pi\gamma^{ij}\right),\label{V_ij}
\end{equation}
and
\begin{eqnarray}
Q^{ij}: & = & \frac{2N}{\gamma}\left(\pi_{k}^{i}\pi^{kj}-\frac{\pi\pi^{ij}}{D-2}\right)-\frac{N}{\gamma}\gamma^{ij}\left(\pi_{kl}^{2}-\frac{\pi^{2}}{D-2}\right)\nonumber \\
 &  & -\frac{1}{\sqrt{\gamma}}D_{k}(N^{k}\pi^{ij})+\frac{2}{\sqrt{\gamma}}\pi^{k(i}D_{k}N^{j)},\label{Q_ij}
\end{eqnarray}
and $Q:=\gamma_{ij}Q^{ij}$. Equation (\ref{esasintegral}) is our
main result: given a solution, that is an initial data, one an compute
this integral which measures the deviation from stationarity. We can
also write (\ref{esasintegral}) in terms of $\gamma_{ij}$ and the
extrinsic curvature $K_{ij}$ . For this purpose all one needs to
do is to rewrite $V^{ij}$ and $Q^{ij}$ in terms of these variables.
They are given as 
\begin{equation}
V^{ij}=2NK^{ij},
\end{equation}
and
\begin{eqnarray}
Q^{ij}: & = & 2N\left(K_{k}^{i}K^{kj}-KK^{ij}\right)-N\gamma^{ij}\left(K_{kl}^{2}-K^{2}\right)\nonumber \\
 &  & -D_{k}(N^{k}K^{ij})+\gamma^{ij}D_{k}(N^{k}K)+2K^{k(i}D_{k}N^{j)}-2KD^{(i}N^{j)}.
\end{eqnarray}
Up to now we have not made a choice of gauge or coordinates. Let us
now choose the Gaussian normal coordinates ( $N=1$, $N^{i}=0$) on
$\Sigma$ for which the integral reads 
\begin{eqnarray}
\text{\ensuremath{\mathscr{I}}}({\cal {N}})= &  & \intop_{\Sigma}dV\Bigg\{\frac{4}{\gamma}\left(|D_{m}\pi^{ij}|^{2}-\frac{D-3}{\left(D-2\right)^{2}}|D_{m}{Tr}(\pi)|^{2}\right)\nonumber \\
 &  & +^{\varSigma}R_{ij}^{2}+\frac{4}{\gamma}{}^{\varSigma}R_{ij}\pi^{ik}\pi_{k}^{j}-\frac{4}{(D-2)\gamma}{}^{\varSigma}R_{ij}\pi^{ij}{Tr}(\pi)\nonumber \\
 &  & -\frac{4}{\gamma}\Lambda\left({Tr}(\pi^{2})-\frac{1}{(D-2)}({Tr}(\pi))^{2})\right)+\frac{D-7}{\gamma^{2}}\left({Tr}(\pi^{2})-\frac{1}{D-2}({Tr}(\pi))^{2}\right)^{2}\nonumber \\
 &  & +\frac{4}{\gamma^{2}}\left({Tr}(\pi^{4})-\frac{2}{D-2}{Tr}(\pi){Tr}(\pi^{3})+\frac{1}{(D-2)^{2}}({Tr}(\pi))^{2}{Tr}(\pi^{2})\right)\Bigg\},
\end{eqnarray}
where ${Tr}(\pi):=\gamma_{ij}\pi^{ij}$ and ${Tr}(\pi^{2}):=\pi^{ij}\pi_{ij}$
and so on. In terms of the extrinsic curvature, in the Gaussian normal
coordinates, one has 
\begin{eqnarray}
\text{\ensuremath{\mathscr{I}}}({\cal {N}}) & = & \intop_{\Sigma}dV\Bigg\{4|D_{m}K_{ij}|^{2}+{}^{\varSigma}R_{ij}^{2}+4{}^{\varSigma}R_{ij}(K^{ik}K_{k}^{j}-KK^{ij})\nonumber \\
 &  & +4\Lambda\left(K^{2}-K_{ij}^{2}\right)+4K_{ij}K^{jl}K_{lm}K^{mi}-8KK_{ij}K^{jl}K_{l}^{i}\\
 &  & -2(D-9)K^{2}K_{ij}^{2}+(D-7)\left((K_{ij}^{2})^{2}+K^{4}\right)\Bigg\}.\nonumber 
\end{eqnarray}
For a physically meaningful solution whose ADM mass and angular momenta
are finite for the asymptotically flat case, or in the case of $\Lambda\neq0$
whose Abbott-Deser \cite{AD} charges are finite, this quantity is
expected to be finite and represents the non-stationary part of the
total energy by construction. Observe that while the ADM momentum
($P_{i}=\ointctrclockwiseop_{\partial\Sigma}K_{ij}dS^{j}$) and angular
momenta ($J^{jk}=\ointctrclockwiseop_{\partial\Sigma}(x^{j}K^{km}-x^{k}K^{jm})dS_{m}$)
are linear in the extrinsic curvature given as integrals over the
boundary, $\text{\ensuremath{\mathscr{I}}}({\cal {N}})$ has quadratic,
cubic and quartic terms in the extrinsic curvature in the bulk integral. 

Before we lay out the details of the above discussion, let us note
that our final formula (\ref{esasintegral}) can be reduced in various
ways depending on the physical problem or the numerical integration
scheme: for example, one can choose the maximal slicing gauge for
which ${Tr}(\pi)=K=0$. If the problem permits time-symmetric initial
data $\pi^{ij}=K^{ij}=0$, then in this restricted case, $V^{ij}=Q^{ij}=0$,
and the integral (\ref{esasintegral}) reduces to 
\begin{eqnarray*}
 &  & \text{\ensuremath{\mathscr{I}}}({\cal {N}})=\intop_{\Sigma}dV\Bigl({}^{\varSigma}R_{ij}^{2}N^{2}+(D_{i}D_{j}N)^{2}-2{}^{\varSigma}R^{ij}ND_{i}D_{j}N+2{}^{\varSigma}RN\triangle N\\
 &  & ~~~~~~~~~~~~~~+4D_{m}D_{i}N_{j}D^{m}D^{(i}N^{j)}+(D-3)\triangle N\triangle N\Bigr).
\end{eqnarray*}
Let us go back to (\ref{bizimdenk}) which was the defining relation
of the invariant and try to write it as a boundary integral over the
boundary of the hypersurface $\Sigma$. Then one has 
\begin{equation}
\text{\ensuremath{\mathscr{I}}}({\cal {N}})=\intop_{\Sigma}dV\thinspace\widetilde{{\cal {P}}}^{*}({\cal {N}})\cdot\widetilde{{\cal {P}}}^{*}({\cal {N}})=\intop_{\Sigma}dV\thinspace{\cal {N}}\cdot\widetilde{{\cal {P}}}\circ\widetilde{{\cal {P}}}^{*}({\cal {N}})+\ointop_{\partial\Sigma}dS\thinspace n^{k}B_{k},
\end{equation}
which requires $\widetilde{{\cal {P}}}\circ\widetilde{{\cal {P}}}^{*}({\cal {N}})=0$.
This the approximate KID equation introduced by Dain \cite{Dain}
and $B_{k}$ is the boundary term to be found below. Note that our
bulk integral (\ref{esasintegral}) is more general and does not assume
the existence of approximate symmetries.

\section{{\normalsize{}DETAILS OF THE CONSTRUCTION IN $D$ DIMENSIONS}}

\subsection{Boundary Integral}

The importance of the Einstein constraints (\ref{Einstein_c}) cannot
be overstated: clearly the initial data is not arbitrary, one must
solve these equations to feed the evolution equations; but, as importantly,
the constraints also determine the evolution equations and they are
related to the symmetries of the spacetime in a rather intricate way
as we have seen above. One can consider the constraints (\ref{Einstein_c})
as the kernel of a map $\Phi$ 
\begin{equation}
\Phi:\mathscr{M}_{2}\times{\cal {S}}_{2}^{*}\rightarrow{\cal {C}}^{*}\times{\cal {X}}^{*},
\end{equation}
where $\text{\ensuremath{\mathscr{M}}}_{2}$ denotes the space of
the Riemannian metrics and ${\cal {S}}_{2}^{*}$ denotes the space
of symmetric rank-2 tensor densities, ${\cal {C}}^{*}$ denotes the
space of scalar function densities and ${\cal {X}}^{*}$ the space
of vector field densities on the hypersurface $\Sigma$. We can express
the constraint map explicitly as 
\begin{equation}
\Phi\begin{pmatrix}\gamma_{ij}\\
\pi^{ij}
\end{pmatrix}=\begin{pmatrix}\sqrt{\gamma}\left(2\Lambda-^{\varSigma}R\right)+\gamma^{-1/2}\left(\pi_{ij}^{2}-\frac{\pi^{2}}{D-2}\right)\\
-2\gamma_{ki}D_{j}\pi^{kj}
\end{pmatrix},
\end{equation}
whose linearization can be found to be 
\begin{equation}
D\Phi\begin{pmatrix}h_{ij}\\
p^{ij}
\end{pmatrix}=\begin{pmatrix}\sqrt{\gamma}\left(^{\varSigma}R{}^{ij}-D^{i}D^{j}+\gamma^{ij}\triangle\right)h_{ij}\\
\frac{1}{\sqrt{\gamma}}\left(\gamma^{ij}\left(\frac{\pi^{2}}{D-2}-\pi_{ij}^{2}\right)+2\left(\pi^{ik}\pi_{k}^{j}-\frac{\pi^{ij}\pi}{D-2}\right)\right)h_{ij}\\
+\frac{2}{\sqrt{\gamma}}\left(\pi_{ij}-\frac{\pi\gamma_{ij}}{D-2}\right)p^{ij}\\
\\
\left(\pi^{ij}D_{k}-2\delta_{k}^{(i}\pi^{j)l}D_{l}\right)h_{ij}-2\gamma_{k(i}D_{j)}p^{ij}
\end{pmatrix}.
\end{equation}
We can define a $2\times2$ matrix as 
\begin{equation}
D\Phi:=\begin{pmatrix}\sqrt{\gamma}\left(^{\varSigma}R{}^{ij}-D^{i}D^{j}+\gamma^{ij}\triangle\right) & \frac{2}{\sqrt{\gamma}}\left(\pi_{ij}-\frac{\pi\gamma_{ij}}{D-2}\right)\\
+\frac{1}{\sqrt{\gamma}}\left(\gamma^{ij}\left(\frac{\pi^{2}}{D-2}-\pi_{ij}^{2}\right)+2\left(\pi^{ik}\pi_{k}^{j}-\frac{\pi^{ij}\pi}{D-2}\right)\right)\\
\\
\pi^{ij}D_{k}-2\delta_{k}^{(i}\pi^{j)l}D_{l} & -2\gamma_{k(i}D_{j)}
\end{pmatrix},
\end{equation}
such that
\begin{equation}
D\Phi\begin{pmatrix}h_{ij}\\
p^{ij}
\end{pmatrix}=D\Phi\circ\begin{pmatrix}h_{ij}\\
p^{ij}
\end{pmatrix}.
\end{equation}
Defining \cite{Bartnik}
\begin{equation}
\widetilde{{\cal {P}}}:=D\Phi\circ\begin{pmatrix}\gamma^{-1/2} & 0\\
0 & -D^{m}
\end{pmatrix},
\end{equation}
one finds
\begin{equation}
\widetilde{{\cal {P}}}:=\begin{pmatrix}^{\varSigma}R{}^{ij}-D^{i}D^{j}+\gamma^{ij}\triangle & \frac{2}{\sqrt{\gamma}}\left(\frac{\pi\gamma_{ij}}{D-2}-\pi_{ij}\right)D^{m}\\
+\frac{1}{\gamma}\left(\gamma^{ij}\left(\frac{\pi^{2}}{D-2}-\pi_{ij}^{2}\right)+2\left(\pi^{ik}\pi_{k}^{j}-\frac{\pi^{ij}\pi}{D-2}\right)\right)\\
\\
\frac{1}{\sqrt{\gamma}}\left(\pi^{ij}D_{k}-2\delta_{k}^{(i}\pi^{j)l}D_{l}\right) & 2\gamma_{k(i}D_{j)}D^{m}
\end{pmatrix},\label{ptildeoperator}
\end{equation}
which is a map as
\begin{equation}
\widetilde{{\cal {P}}}:{\cal {S}}_{2}\times{\cal {S}}_{1,2}\rightarrow{\cal {C}}\times{\cal {X}},
\end{equation}
where ${\cal {S}}_{2}$ denotes the space of covariant rank-2 tensors,
${\cal {S}}_{1,2}$ denotes the space of covariant rank-3 tensors
which are symmetric in last two indices, ${\cal {C}}$ denotes the
space of scalar function and ${\cal {X}}$ the space of vector fields
on the hypersurface $\Sigma$.

The formal adjoint of $\widetilde{{\cal {P}}}$-operator was defined
in (\ref{formaladjoint_ptilde}) via the (\ref{formaladjointoflinearizedmap_pi})
and it is a map of the form
\begin{equation}
\widetilde{{\cal {P}}}^{*}:{\cal {C}}\times{\cal {X}}\rightarrow{\cal {S}}_{2}\times{\cal {S}}_{1,2}.
\end{equation}
 Working out the details, one arrives at 
\begin{equation}
\widetilde{{\cal {P}}}^{*}\begin{pmatrix}N\\
N^{k}
\end{pmatrix}=\begin{pmatrix}N^{\varSigma}R{}^{ij}-D^{i}D^{j}N+\gamma^{ij}\triangle N+Q^{ij}\\
D_{m}\left(2D_{(i}N_{j)}+V_{ij}\right)
\end{pmatrix},\label{ptildeadjoint}
\end{equation}
where $V^{ij}$ and $Q^{ij}$ were given (\ref{V_ij},\ref{Q_ij})
respectively. We have used this expression in the previous section
to find the bulk integral of the non-stationary energy. Now let us
use this operator and its adjoint to find an expression on the boundary.
For this purpose we need the following identity:

\begin{equation}
\intop_{\Sigma}dV\thinspace\begin{pmatrix}N\\
N^{k}
\end{pmatrix}\cdot{\cal \widetilde{{\cal {P}}}}\begin{pmatrix}s_{ij}\\
s_{kij}
\end{pmatrix}=\intop_{\Sigma}dV\begin{pmatrix}s_{ij}\\
s_{kij}
\end{pmatrix}\cdot\widetilde{{\cal {P}}}^{*}\begin{pmatrix}N\\
N^{k}
\end{pmatrix}+\ointop_{\partial\Sigma}dS\thinspace n^{k}{\cal {B}}{}_{k},\label{integrall}
\end{equation}
with generic $s_{ij}\in{\cal {S}}_{2}$ and $s_{kij}\in{\cal {S}}_{1,2}$.
After making use of (\ref{ptildeoperator}) and (\ref{ptildeadjoint}),
a slightly cumbersome computation yields the boundary term:
\begin{eqnarray}
 &  & {\cal {B}}{}_{k}=s_{kj}D^{j}N-ND^{j}s_{kj}+ND_{k}s-sD_{k}N+2N^{i}D^{j}s_{jki}-2s_{kij}D^{i}N^{j}\nonumber \\
 &  & ~~~~~~~~~~+\frac{2N}{\sqrt{\gamma}}\left(\frac{\pi}{D-2}s_{kj}\thinspace^{j}-s_{kij}\pi^{ij}\right)+\frac{1}{\sqrt{\gamma}}\left(\pi^{ij}s_{ij}N_{k}-2s_{ij}N^{i}\pi_{k}^{j}\right),\label{B_k}
\end{eqnarray}
where $s=\gamma^{ij}s_{ij}$. Let us now assume a particular $s_{ij}$
and a particular $s_{kij}$ such that
\begin{equation}
\begin{pmatrix}s_{ij}\\
s_{kij}
\end{pmatrix}:=\widetilde{{\cal {P}}}^{*}\begin{pmatrix}N\\
N^{k}
\end{pmatrix},
\end{equation}
which yields
\begin{equation}
{\cal \widetilde{{\cal {P}}}}\begin{pmatrix}s_{ij}\\
s_{kij}
\end{pmatrix}={\cal \widetilde{{\cal {P}}}}\circ\widetilde{{\cal {P}}}^{*}\begin{pmatrix}N\\
N^{k}
\end{pmatrix}.
\end{equation}
Then (\ref{integrall}) becomes
\begin{equation}
\intop_{\Sigma}dV\thinspace\begin{pmatrix}N\\
N^{k}
\end{pmatrix}\cdot{\cal \widetilde{{\cal {P}}}}\circ\widetilde{{\cal {P}}}^{*}\begin{pmatrix}N\\
N^{k}
\end{pmatrix}=\text{\ensuremath{\mathscr{I}}}({\cal {N}})+\ointop_{\partial\Sigma}dS\thinspace n^{k}{\cal {B}}{}_{k},\label{integrallreduced}
\end{equation}
where ${\cal {B}}{}_{k}$ given in (\ref{B_k}) must be evaluated
with 
\begin{equation}
s_{ij}=N{}^{\varSigma}R_{ij}-D_{i}D_{j}N+\gamma_{ij}\triangle N+Q_{ij}
\end{equation}
and
\begin{equation}
s_{kij}=D_{k}\left(2D_{(i}N_{j)}+V_{ij}\right).
\end{equation}
Equation (\ref{integrallreduced}) shows that generically $\text{\ensuremath{\mathscr{I}}}({\cal {N}})$
cannot be written as an integral on the boundary of the hypersurface
unless ${\cal \widetilde{{\cal {P}}}}\circ\widetilde{{\cal {P}}}^{*}({\cal {N}})=0$.
In that case, the invariant reduces to 
\begin{equation}
\text{\ensuremath{\mathscr{I}}}({\cal {N}})=-\ointop_{\partial\Sigma}dS\thinspace n^{k}{\cal {B}}{}_{k}.
\end{equation}
Explicit computation shows that one has 
\begin{eqnarray}
 &  & {\cal {B}}{}_{k}=\frac{N^{2}}{2}D_{k}{}^{\Sigma}R+N{}^{\Sigma}R_{kj}D^{j}N-D_{k}D_{j}ND^{j}N-(D-3)D_{k}N\triangle N+(D-2)ND_{k}\triangle N\nonumber \\
 &  & ~~~~~~~~~~+4N^{i}\triangle D_{(k}N_{i)}-4D_{k}D_{(i}N_{j)}D^{(i}N^{j)}+b_{k},\label{B_k-1}
\end{eqnarray}
where
\begin{eqnarray}
 &  & b_{k}:=Q_{kj}D^{j}N-ND^{j}Q_{kj}+ND_{k}Q-QD_{k}N+2N^{i}\triangle V_{ki}-2D_{k}V_{ij}D^{i}N^{j}\nonumber \\
 &  & ~~~~~~~~~+\frac{1}{\sqrt{\gamma}}\frac{2N\pi}{D-2}\left(2D_{k}D_{i}N^{i}+D_{k}V\right)-\frac{2N\pi^{ij}}{\sqrt{\gamma}}\left(2D_{k}D_{i}N_{j}+D_{k}V_{ij}\right)\nonumber \\
 &  & ~~~~~~~~~+\frac{1}{\sqrt{\gamma}}\left(\pi^{ij}N_{k}-2N^{i}\pi_{k}^{j}\right)\left(N^{\varSigma}R_{ij}-D_{i}D_{j}N+\gamma_{ij}\triangle N+Q_{ij}\right).\label{b_k}
\end{eqnarray}
In the Gaussian normal coordinates the boundary integral reads
\begin{equation}
\text{\ensuremath{\mathscr{I}}}({\cal {N}})=\ointop_{\partial\Sigma}dS\thinspace n^{k}\left((D-\frac{5}{2})D_{k}K_{ij}^{2}+(\frac{7}{2}-D)D_{k}K^{2}+2K^{lj}D_{j}K_{lk}\right).
\end{equation}
Another physically relevant case is the time symmetric asymptotically
flat case for which the boundary integral reduces to 
\begin{eqnarray*}
 &  & \text{\ensuremath{\mathscr{I}}}({\cal {N}})=\ointop_{\partial\Sigma}dS\thinspace n^{k}\Bigl(D_{k}D_{j}ND^{j}N+(D-3)D_{k}N\triangle N-(D-2)ND_{k}\triangle N\\
 &  & ~~~~~~~~~~~~-4N^{i}\triangle D_{(k}N_{i)}+4D_{k}D_{(i}N_{j)}D^{(i}N^{j)}\Bigr).
\end{eqnarray*}
In the most general form $N$ and $N^{i}$ should satisfy the fourth
order equations ${\cal \widetilde{{\cal {P}}}}\circ\widetilde{{\cal {P}}}^{*}({\cal {N}})=0$
which explicitly read
\begin{equation}
\widetilde{{\cal {P}}}\text{\textopenbullet}\widetilde{{\cal {P}}}^{*}\begin{pmatrix}N\\
N^{i}
\end{pmatrix}=\begin{pmatrix}(D-2)\triangle\triangle N-{}^{\Sigma}R{}_{ij}D^{i}D^{j}N+N\left(\frac{1}{2}\triangle{}^{\Sigma}R+{}^{\Sigma}R{}_{ij}^{2}\right)\\
+2\,{}^{\Sigma}R\triangle N+\frac{3}{2}D_{i}{}^{\Sigma}RD^{i}N+Y\\
\\
4D^{j}\triangle D_{(k}N_{j)}+Y_{k}
\end{pmatrix}=0,\label{PcompositionPadjointDdimensions}
\end{equation}
where
\begin{eqnarray}
 &  & Y:=^{\Sigma}R{}^{ij}Q_{ij}-D^{i}D^{j}Q_{ij}+\triangle Q+\frac{2}{\sqrt{\gamma}}\left(\frac{\pi\gamma^{ij}}{D-2}-\pi^{ij}\right)\triangle(2D_{i}N_{j}+V_{ij})\label{YY}\\
 &  & ~~~+\left(\frac{2}{\gamma}(\pi^{ik}\pi_{k}^{j}-\frac{\pi\pi^{ij}}{D-2})-\frac{\gamma^{ij}}{\gamma}(\pi_{kl}^{2}-\frac{\pi^{2}}{D-2})\right)\left(N^{\varSigma}R_{ij}-D_{i}D_{j}N+\gamma_{ij}\triangle N+Q_{ij}\right)\nonumber 
\end{eqnarray}
and 
\begin{equation}
Y_{k}:=\frac{1}{\sqrt{\gamma}}\left(\pi^{ij}D_{k}-2\delta_{k}^{i}\pi^{jl}D_{l}\right)\left(N^{\varSigma}R_{ij}-D_{i}D_{j}N+\gamma_{ij}\triangle N+Q_{ij}\right)+2D^{i}\triangle V_{ik}.
\end{equation}

\subsection{{\normalsize{}{}{}The Approximate KID equation in D dimensions}}

Following the $D=4$ discussion of Dain \cite{Dain} let us now study
the approximate KID equation (\ref{PcompositionPadjointDdimensions})
in $D$ dimensions. It is easy to see that it is a fourth order elliptic
operator for $D>2$. This follows by computing the leading symbol:
for this purpose let us consider the higher order derivative terms
and set $D_{i}=\zeta_{i}$ and $\left|\zeta\right|^{2}=\zeta^{i}\zeta_{i}$.
Using (\ref{PcompositionPadjointDdimensions}), the leading symbol
of operator reads 
\begin{equation}
\sigma\left[{\cal \widetilde{{P}}}\text{\textopenbullet}\widetilde{{\cal {P}}}^{*}\right]\left(\zeta\right)\begin{pmatrix}N\\
N_{i}
\end{pmatrix}=\begin{pmatrix}(D-2)\left|\zeta\right|^{4}N\\
4\left|\zeta\right|^{2}\zeta^{j}\zeta_{(k}N_{j)}
\end{pmatrix}.
\end{equation}
For a non-zero covector $\zeta$, if $\sigma$ is an isomorphism (here
a vector bundle isomorphism), then the operator is elliptic. For the
first component, this requires $D\neq2$ and for the second component
contraction with $\zeta^{k}$ yields 
\begin{equation}
\left|\zeta\right|^{4}\zeta^{k}N_{k}=0.
\end{equation}
Assuming $D\neq2$ one has $\zeta^{k}N_{k}=0$. Inserting it back
in the second component one obtains 
\begin{equation}
\left|\zeta\right|^{4}N_{k}=0,
\end{equation}
so for $\left|\zeta\right|^{2}\neq0$, the leading symbol is injective
and the operator ${\cal \widetilde{{P}}}\text{\textopenbullet}\widetilde{{\cal {P}}}^{*}$
is elliptic for $D>2$.

\subsection{{\normalsize{}{}{}Asymptotically Flat Spaces}}

Consider the initial data set $\left(\Sigma,\gamma_{ij},\pi^{ij}\right)$
for the vacuum Einstein field equations with $n>1$ asymptotically
Euclidean ends: this is to avoid bulk simplicity and allow black holes.
There exists a compact set ${\cal {B}}$ such that ${\Sigma}\setminus{\cal {B}}=\sum_{k=1}^{n}{\Sigma}_{(k)}$
, where ${\Sigma}_{(k)}$, $k=1,...,n$ are open sets diffeomorphic
to the complement of a closed ball in $\mathbb{R}^{D-1}$. Each asymptotic
end ${\Sigma}_{(k)}$ admits asymptotically Cartesian coordinates.
We consider the following decay assumptions, for $D>3$, which are
consistent with finite ADM mass and momenta:

\begin{equation}
\gamma_{ij}=\delta_{ij}+o(\left|x\right|^{(3-D)/2}),\label{decayassumptionh}
\end{equation}
\begin{equation}
\pi^{ij}=o(\left|x\right|^{(1-D)/2}),\label{decayassumptionk}
\end{equation}
where $\delta_{ij}=(++...+)$. Note that $\delta_{ij}=\mathcal{O}(1)$
and beware of the small $o$and and the big $\mathcal{O}$ notation.
One can compute the the following decay behavior for the Christoffel
connection

\begin{equation}
^{\Sigma}\Gamma_{ij}^{k}=o(\left|x\right|^{(1-D)/2}),
\end{equation}
and the curvatures 
\begin{equation}
^{\Sigma}R^{k}\thinspace_{lmn}=o(\left|x\right|^{-(1+D)/2}),~~~~~~~{}^{\Sigma}R{}_{ij}=o(\left|x\right|^{-(1+D)/2}),~~~~~~~~{}^{\Sigma}R=o(\left|x\right|^{-(1+D)/2}).
\end{equation}

\subsection{{\normalsize{}{}{}KIDs in $D$ Dimensions}}

Let $\left(\Sigma,\gamma_{ij},\pi^{ij}\right)$ denote a smooth vacuum
initial data set satisfying the decay assumptions (\ref{decayassumptionh},
\ref{decayassumptionk}). Let $N,N^{i}$ be a smooth scalar field
and a vector field on $\Sigma$ satisfying the KID equations. Then
generalizing the $D=4$ result of \cite{BeiChr96}, the behavior of
all the possible solutions were given in \cite{Chrusciel:2005pb}
which we quote here. 
\begin{enumerate}
\item There exits an antisymmetric tensor field $\mathscr{\omega}_{\mu\nu}$,
such that 
\begin{equation}
N-\omega_{0i}x^{i}=o(\left|x\right|^{(5-D)/2}),~~~~~~~~~~~~N^{i}-\omega^{i}\thinspace_{j}x^{j}=o(\left|x\right|^{(5-D)/2}).
\end{equation}
\item If $\omega_{\mu\nu}=0$, then there exists a vector field ${\cal {U}}^{\mu}$,
such that 
\begin{equation}
N-{\cal {U}}^{0}=o(\left|x\right|^{(3-D)/2}),~~~~~~~~~~~~~~~N^{i}-{\cal {U}}^{i}=o(\left|x\right|^{(3-D)/2}).\label{stationaryKIDs}
\end{equation}
\item If $\omega_{\mu\nu}=0={\cal {U}}^{\mu}$ then one has the trivial
solution $N=0=N^{i}$ . Both $\mathscr{\omega}_{\mu\nu}$ and ${\cal {U}}^{\mu}$
are constants in the sense that they are $\mathcal{O}(1)$ whenever
they don't vanish. 
\end{enumerate}
Case $1$ above corresponds to the rotational Killing vectors while
case $2$ corresponds to the translational ones we shall employ the
latter.

We explained in section $\text{\mbox{II}}$ that solutions of the
$D\Phi^{*}(N,N^{i})=0$ yield spacetime Killing vectors. It is not
difficult to see that the modified equation $\widetilde{{\cal {P}}}^{*}\left(N,N^{i}\right)=0$
yields only the Killing vectors for the case of translational KIDs
(\ref{decayassumptionh},\ref{decayassumptionk}). Here is the proof:
$\widetilde{{\cal {P}}}^{*}\left(N,N^{i}\right)=0$ implies 
\begin{equation}
N^{\Sigma}R{}_{ij}-D_{i}D_{j}N+\gamma_{ij}\triangle N+Q_{ij}=0,
\end{equation}
\begin{equation}
D_{m}\left(2D_{(i}N_{j)}+V_{ij}\right)=0.
\end{equation}
If one assumes $\left(N,N^{i}\right)$ decay as in (\ref{stationaryKIDs})
we have $D_{(i}N_{j)}=o(\left|x\right|^{(1-D)/2})$; and $V_{ij}=o(\left|x\right|^{(1-D)/2})$,
then 
\begin{equation}
2D_{(i}N_{j)}+V_{ij}=o(\left|x\right|^{(1-D)/2})
\end{equation}
vanishes at infinity; and since it is covariantly constant, it must
vanish identically 
\begin{equation}
2D_{(i}N_{j)}+V_{ij}=0.
\end{equation}
Together with the first component of $\widetilde{{\cal {P}}}^{*}\left(N,N^{i}\right)=0$
we get the formal adjoint of the linearized constraint map, namely
$D\Phi^{*}(N,N^{i})=0$. We can conclude that if $\widetilde{{\cal {P}}}^{*}\left(N,N^{i}\right)=0$
then $\left(N,N^{i}\right)$ solve the KID equations.

\subsection{Approximate KIDs in $D$ dimensions}

Generalizing Dain's $D=4$ result, let us search for translational
solutions of the approximate Killing equation \footnote{We work in a given asymptotic end and not the clutter the notation
we do not denote the corresponding index referring to the asymptotic
end. } 
\begin{equation}
\widetilde{{\cal {P}}}\circ\widetilde{{\cal {P}}}^{*}\begin{pmatrix}N\\
N^{i}
\end{pmatrix}=0\label{yenia}
\end{equation}
as a deformation of the KIDs $(X,N^{i})$ in the following form: 
\begin{equation}
N=\lambda\varphi+X,~~~~~~~~~~N^{i}=N^{i},\label{ansatz}
\end{equation}
where the function $\varphi$ is to be found, $\lambda$ is a constant.
KIDs decay as 
\begin{equation}
X-{\cal {U}}^{0}=o(\left|x\right|^{(3-D)/2}),
\end{equation}
\begin{equation}
N^{i}-{\cal {U}}^{i}=o(\left|x\right|^{(3-D)/2}).
\end{equation}
Inserting the ansatz (\ref{ansatz}) into the approximate KID equation
(\ref{yenia}), one gets 
\begin{equation}
\widetilde{{\cal {P}}}\circ\widetilde{{\cal {P}}}^{*}\begin{pmatrix}\varphi\\
0
\end{pmatrix}=-\widetilde{{\cal {P}}}\circ\widetilde{{\cal {P}}}^{*}\begin{pmatrix}X\\
N^{i}
\end{pmatrix}=0,\label{important}
\end{equation}
or more explicitly
\begin{equation}
\widetilde{{\cal {P}}}\circ\widetilde{{\cal {P}}}^{*}\begin{pmatrix}\varphi\\
0
\end{pmatrix}=\begin{pmatrix}(D-2)\triangle\triangle\varphi-{}^{\Sigma}R{}_{ij}D^{i}D^{j}\varphi+\varphi\left(\frac{1}{2}\triangle{}^{\Sigma}R+{}^{\Sigma}R_{ij}^{2}\right)\\
+2^{\Sigma}R\triangle\varphi+\frac{3}{2}D_{i}{}^{\Sigma}RD^{i}\varphi+Y\\
\\
Y_{k}
\end{pmatrix}=0.\label{matrix}
\end{equation}
For such a $\varphi$, the bulk integral (\ref{esasintegral}) becomes

\begin{eqnarray}
 &  & \text{\ensuremath{\mathscr{I}}}({\cal {N}})=\lambda^{2}\intop_{\Sigma}dV\Bigg\{|D_{m}V^{ij}|^{2}+{}^{\varSigma}R_{ij}^{2}\varphi^{2}+(D_{i}D_{j}\varphi)^{2}-2{}^{\varSigma}R^{ij}ND_{i}D_{j}\varphi+2{}^{\varSigma}R\varphi\triangle\varphi\nonumber \\
 &  & ~~~~~~~~~~~~~~~~~+(D-3)\triangle\varphi\triangle\varphi+2Q\triangle\varphi+Q_{ij}^{2}+2{}^{\varSigma}R_{ij}\varphi Q^{ij}-2Q^{ij}D_{i}D_{j}\varphi\Bigr\},\label{eb5}
\end{eqnarray}
where
\begin{equation}
V^{ij}=2\varphi K^{ij},
\end{equation}
and
\begin{equation}
Q^{ij}=2\varphi\left(K_{k}^{i}K^{kj}-KK^{ij}\right)-\varphi\gamma^{ij}\left(K_{kl}^{2}-K^{2}\right).
\end{equation}
The boundary form for the asymptotically flat case follows similarly
\begin{eqnarray}
 &  & \text{\ensuremath{\mathscr{I}}}({\cal {N}})=-\lambda^{2}\ointop_{\partial\Sigma}dS\thinspace n^{k}\Bigl\{-D_{k}D_{j}\varphi D^{j}\varphi-(D-3)D_{k}\varphi\triangle\varphi+(D-2)\varphi D_{k}\triangle\varphi\nonumber \\
 &  & +Q_{kj}D^{j}\varphi-\varphi D^{j}Q_{kj}-2\varphi K^{ij}D_{k}V_{ij}\Bigr\},\label{asdfdsa}
\end{eqnarray}
where we used $K_{kl}^{2}-K^{2}={}^{\Sigma}R=0$ on the boundary. 

\section{{\normalsize{}{}{}Conclusions}}

Using the Hamiltonian form of the Einstein evolution equations as
given by Fischer and Marsden \cite{FisMar72}, we constructed an integral
that measures the non-stationary energy contained in a spacelike hypersurface in $D$ dimensional General Relativity with or without a cosmological constant.  
This integral was previously studied by Dain \cite{Dain} who used
the Einstein constraints but not the evolution equations. The crucial observation
is the following: the critical points of the first order Hamiltonian form of Einstein equations
correspond to the initial data which possess Killing symmetries, a
result first observed by Moncrief \cite{Moncrief}. Hence, our vantage
point is that the failure of an initial data to possess Killing symmetries
is given by the evolution equations, namely non-vanishing of the time derivatives of the spatial metric and the canonical momenta. Then manipulating the evolution equations,
one arrives at the integral (\ref{esasintegral}). Once an initial
data is given, one can compute this integral, which by construction,
vanishes for stationary spacetimes. 

\section{{\normalsize{}{}{}Acknowledgments}}
This work is dedicated to the memory of Rahmi G\"{u}ven (1948-2019) who spent a life in gravity research in a region quite timid about science. 

\section{{\normalsize{}{}{}Appendix: ADM Split of Einstein's Equations in $D$ Dimensions}}

For the sake of completeness let us give here the ADM split of Einstein's
equations and all the relevant tensors. Using the $\left(D-1\right)+1$
dimensional decomposition of the metric given as (\ref{ADMdecompositionofmetric})
we have:
\begin{equation}
g_{00}=-N^{2}+N_{i}N^{i},~\ ~~~g_{0i}=N_{i},~~~~\ g_{ij}=\gamma_{ij},
\end{equation}
and 
\begin{equation}
g^{00}=-\frac{1}{N^{2}},~~~~g^{0i}=\frac{1}{N^{2}}N^{i},~~~~g^{ij}=\gamma^{ij}-\frac{1}{N^{2}}N^{i}N^{j}.
\end{equation}
Let $\Gamma_{\nu\rho}^{\mu}$ denote the Christoffel symbol of the
$D$ dimensional spacetime 
\begin{equation}
\Gamma_{\nu\rho}^{\mu}=\frac{1}{2}g^{\mu\sigma}\left(\partial_{\nu}g_{\rho\sigma}+\partial_{\rho}g_{\nu\sigma}-\partial_{\sigma}g_{\nu\rho}\right)
\end{equation}
and let $^{\Sigma}\Gamma_{ij}^{k}$ denote the Christoffel symbol
of the $D-1$ dimensional hypersurface, which is compatible with the
spatial metric $\gamma_{ij}$ as 
\begin{equation}
^{\Sigma}\Gamma_{ij}^{k}=\frac{1}{2}\gamma^{kp}\left(\partial_{i}\gamma_{jp}+\partial_{j}\gamma_{ip}-\partial_{p}\gamma_{ij}\right).\label{Christoffelofhypersurface}
\end{equation}
Then a simple computation shows that 
\begin{equation}
\Gamma_{00}^{0}=\frac{1}{N}\left(\dot{N}+N^{k}(\partial_{k}N+N^{i}K_{ik})\right)
\end{equation}
and
\begin{equation}
\Gamma_{0i}^{0}=\frac{1}{N}\left(\partial_{i}N+N^{k}K_{ik}\right),~~~~\Gamma_{ij}^{0}=\frac{1}{N}K_{ij},~~~~\Gamma_{ij}^{k}=^{\Sigma}\Gamma_{ij}^{k}-\frac{N^{k}}{N}K_{ij}
\end{equation}
and 
\begin{equation}
\Gamma_{0j}^{i}=-\frac{1}{N}N^{i}\left(\partial_{j}N+K_{kj}N^{k}\right)+NK_{j}\thinspace^{i}+D_{j}N^{i}
\end{equation}
and also
\begin{equation}
\Gamma_{00}^{i}=-\frac{N^{i}}{N}\left(\dot{N}+N^{k}\left(\partial_{k}N+N^{l}K_{kl}\right)\right)+N\left(\partial^{i}N+2N^{k}K_{k}\thinspace^{i}\right)+\dot{N}^{i}+N^{k}D_{k}N^{i}.
\end{equation}

Starting with the definition of the $D$ dimensional Ricci tensor
\begin{equation}
R_{\rho\sigma}=\partial_{\mu}\Gamma_{\rho\sigma}^{\mu}-\partial_{\rho}\Gamma_{\mu\sigma}^{\mu}+\Gamma_{\mu\nu}^{\mu}\Gamma_{\rho\sigma}^{\nu}-\Gamma_{\sigma\nu}^{\mu}\Gamma_{\mu\rho}^{\nu}
\end{equation}
one arrives at

\begin{equation}
R_{ij}={}^{\Sigma}R_{ij}+KK_{ij}-2K_{ik}K_{j}^{k}+\frac{1}{N}\left(\dot{K}_{ij}-N^{k}D_{k}K_{ij}-D_{i}D_{j}N-K_{ki}D_{j}N^{k}-K_{kj}D_{i}N^{k}\right),\label{eq:rij}
\end{equation}
where $^{\Sigma}R_{ij}$ denotes the Ricci tensor of the hypersurface.
The remaining components can also be found to be

\begin{equation}
R_{00}=N^{i}N^{j}R_{ij}-N^{2}K_{ij}K^{ij}+N\left(D_{k}D^{k}N-\dot{K}-N^{k}D_{k}K+2N^{k}D_{m}K_{k}^{m}\right)\label{r00}
\end{equation}
and
\begin{equation}
R_{0i}=N^{j}R_{ij}+N\left(D_{m}K_{i}^{m}-D_{i}K\right).\label{eq:ri0}
\end{equation}
The scalar curvature can be found as

\begin{equation}
R=^{\Sigma}R+K^{2}+K_{ij}K^{ij}+\frac{2}{N}\left(\dot{K}-D_{k}D^{k}N-N^{k}D_{k}K\right)\label{r}
\end{equation}
Under the above splitting the cosmological Einstein equations 
\begin{equation}
R_{\mu\nu}-\frac{1}{2}g_{\mu\nu}R+\varLambda g_{\mu\nu}=\kappa T_{\mu\nu}
\end{equation}
split in to constraints and evolution equations in local coordinates.
The momentum constraints read

\begin{equation}
N\left(D_{k}K_{i}^{k}-D_{i}K\right)-\kappa\left(T_{0i}-N^{j}T_{ij}\right)=0,
\end{equation}
via the Hamiltonian constraint becomes

\begin{equation}
N^{2}\left(^{\Sigma}R+K^{2}-K_{ij}^{2}-2\Lambda\right)-2\kappa\left(T_{00}+N^{i}N^{j}T_{ij}-2N^{i}T_{0i}\right)=0.
\end{equation}
On the other hand the evolution equations for the metric and the extrinsic
curvature become
\begin{equation}
\frac{\partial}{\partial t}\gamma_{ij}=2NK_{ij}+D_{i}N_{j}+D_{j}N_{i},
\end{equation}
\begin{equation}
\frac{\partial}{\partial t}K_{ij}=N\left(R_{ij}-{}^{\Sigma}R_{ij}-KK_{ij}+2K_{ik}K_{j}^{k}\right)+\mathscr{L}_{\overrightarrow{N}}K_{ij}+D_{i}D_{j}N,
\end{equation}
where $\text{\ensuremath{\mathscr{L}}}_{\overrightarrow{N}}$ is the
Lie derivative along the shift vector.


\begin{thebibliography}{10}
\bibitem{Dain}S. Dain, A New Geometric Invariant on Initial Data
for the Einstein Equations, Phys. Rev. Lett. \textbf{93}, 23, 231101
(2004).

\bibitem{ADM} R.~Arnowitt, S.~Deser and C.~Misner, The
Dynamics of General Relativity, Phys. \ Rev.\ \textbf{116},
1322 (1959); \textbf{117}, 1595 (1960); in \textit{{Gravitation:
An Introduction to Current Research}}, ed L. Witten (Wiley, New York,
1962).

\bibitem{Kroon}J. A. V. Kroon and J. L. Williams, Dain's invariant
on non-time symmetric initial data sets, Class. Quantum Grav. \textbf{34},
12, 125013, (2017)

\bibitem{merger} B.~P.~Abbott \textit{et al.} {[}LIGO Scientific
and Virgo Collaborations{]}, Observation of Gravitational Waves
from a Binary Black Hole Merger, Phys.\ Rev.\ Lett.\ \textbf{116},
no. 6, 061102 (2016).

\bibitem{Moncrief}V. Moncrief, Spacetime symmetries and linearization
stability of the Einstein equations. I, J. Math. Phys. \textbf{16}
, 493-498 (1975).

\bibitem{Beig} R.~Beig \& P.~T. Chru\'{s}ciel, Killing initial
data, Class. Quantum Grav. \textbf{14}, A83 (1997).

\bibitem{Bartnik}R. Bartnik, Phase space for the Einstein equations,
Communications in Analysis and Geometry \textbf{13}, 845 (2005).

\bibitem{FisMar72} A.~E. Fischer \& J.~E. Marsden, 
The Einstein evolution equations as a first-order quasi-linear symmetric
hyperbolic system. I., \newblock Comm. Math. Phys. \textbf{28},
1 (1972).

\bibitem{BeiChr96} R.~Beig \& P.~T. Chru\'{s}ciel, 
Killing vectors in asymptotically flat spacetimes. {I}. {A}symptotically
translational Killing vectors and rigid positive energy theorem,
\newblock J. Math. Phys. \textbf{37}, 1939 (1996).

\bibitem{Chrusciel:2005pb} P.~T.~Chrusciel and D.~Maerten, Killing
vectors in asymptotically flat space-times. II. Asymptotically translational
Killing vectors and the rigid positive energy theorem in higher dimensions,
J.\ Math.\ Phys.\ \textbf{47}, 022502 (2006).

\bibitem{DeWitt} B.~S.~DeWitt, Quantum Theory of Gravity. 1. The
Canonical Theory, Phys.\ Rev.\ \textbf{160}, 1113 (1967).


\bibitem{Fischer-Marsden}A. E. Fischer and J. E. Marsden, Linearization
stability of the Einstein equations, Bull. Amer. Math. Soc., \textbf{79},
997-1003 (1973).


\bibitem{AD} L. F. Abbott and S. Deser, Stability of gravity with
a cosmological constant, Nucl. Phys. B \textbf{195}, 76 (1982).


\end{thebibliography}
\end{document}